\documentclass[conference]{IEEEtran}
\IEEEoverridecommandlockouts

\usepackage{cite}
\usepackage{amsmath,amssymb,amsfonts}
\usepackage{algorithmic}
\usepackage{graphicx}
\usepackage{textcomp}
\usepackage{xcolor}
\usepackage[hidelinks]{hyperref}

\def\BibTeX{{\rm B\kern-.05em{\sc i\kern-.025em b}\kern-.08em
    T\kern-.1667em\lower.7ex\hbox{E}\kern-.125emX}}
\begin{document}


\title{Caliper-in-the-Loop: Black-Box Optimization for Hyperledger Fabric Performance Tuning}

\author{
\IEEEauthorblockN{
  Yash Madhwal\IEEEauthorrefmark{1},
  Arseny Bolotnikov\IEEEauthorrefmark{2}, 
  Mark Prikhno\IEEEauthorrefmark{3}, 
  Irina Lebedeva\IEEEauthorrefmark{1},
  Ivan Laishevskiy\IEEEauthorrefmark{4}, \\ 
  Vladimir Gorgadze\IEEEauthorrefmark{5}\IEEEauthorrefmark{6}, 
  Artem Barger\IEEEauthorrefmark{7}, 
  Yury Yanovich\IEEEauthorrefmark{1}
}
\IEEEauthorblockA{\IEEEauthorrefmark{1}%
  Skolkovo Institute of Science and Technology, Moscow, Russia \\
}
 \IEEEauthorblockA{\IEEEauthorrefmark{2}%
  HSE University, Moscow, Russia \\
}
 \IEEEauthorblockA{\IEEEauthorrefmark{3}%
  Sberbank PJSC, Moscow, Russia}
  \IEEEauthorblockA{\IEEEauthorrefmark{4}%
  Mid Hope Technologies, Moscow, Russia} 
\IEEEauthorblockA{\IEEEauthorrefmark{5}%
  Moscow Institute of Physics and Technology, Moscow, Russia \\
}
\IEEEauthorblockA{\IEEEauthorrefmark{6}%
  IDEAS: Inter-Disciplinary \& Advanced Studies Center, Moscow, Russia \\
}
\IEEEauthorblockA{\IEEEauthorrefmark{7}%
  University of Jyväskylä, Jyväskylä, Finland \\
}
}

\maketitle

\begin{abstract}
Hyperledger Fabric performance depends on a large number of interacting configuration parameters, making manual tuning expensive and unreliable. We study automated throughput tuning of HLF by modeling benchmarking as a noisy black-box function and applying Bayesian optimization (BO) with dimensionality-reduction (DR) strategies. We implement an end-to-end closed-loop pipeline that (1) deploys HLF with a candidate configuration, (2) benchmarks it with Hyperledger Caliper, and (3) proposes the next configuration based on observed throughput. The search space is derived from HLF configuration files and contains 317 dimensions. To improve initial coverage, we use a Latin Hypercube design and then perform iterative optimization with a fixed evaluation budget per trial. In a cloud testbed, we evaluate 16 acquisition-function/DR combinations and a random-search baseline, executed in parallel on 17 virtual machines. Several BO+DR variants yield measurable throughput improvements relative to the first evaluated configuration; the best-performing approach (DYCORS-PCA) achieves a 12\% TPS increase, and MPI-REMBO achieves a 9\% TPS increase under the tested workload. The results indicate that BO combined with DR can be effective for HLF tuning in very high-dimensional spaces, while also exposing practical limitations driven by measurement noise and the sensitivity of conclusions to the chosen baseline and evaluation protocol.
\end{abstract}

\begin{IEEEkeywords}
Benchmarking, Bayesian optimization, Dimensionality reduction, Hyperledger Fabric, Tuning.
\end{IEEEkeywords}

\section{Introduction}\label{sec:intro}

Hyperledger Fabric (HLF) is a widely adopted framework for deploying permissioned blockchain networks in enterprise environments, where governance, access control, and operational policies must be enforced among known organizations~\cite{antwi2021case,shammar2022attribute,li2020survey}. Its modular architecture, separating identity management, endorsement, ordering, and state management, supports flexible deployments and fine-grained configuration\cite{fabric_eurosys18,fabric_ordering_service_docs}. However, this flexibility introduces a substantial configuration burden: HLF exposes hundreds of configurable parameters spanning peer and orderer settings, channel and batching policies, and state database options~\cite{li2023athena,baset2018hands}. These parameters interact in complex ways, and previous studies show that performance is sensitive to multiple configuration choices, making manual adjustment expensive, time-consuming, and difficult to reproduce~\cite{thakkar2018fabric,wang2020fabric_bottleneck}.

In enterprise deployments, different organizations or consortia run their own independent HLF blockchains, each with tailored configurations. Unlike public blockchains that operate a single main network, these permissioned blockchains are isolated deployments designed for specific business needs. Currently, engineers launching a new HLF network must tune its configuration from scratch, which represents significant operational overhead and requires specialized expertise. However, engineering teams that deploy multiple blockchains over time can accumulate valuable configuration knowledge from their launch history. Our work investigates whether optimization knowledge gained from tuning one blockchain deployment can accelerate the optimization of others--a form of transfer learning for blockchain configuration. While this paper focuses on single-network optimization as a necessary first step, our methodology is designed with multi-deployment knowledge transfer as an explicit future goal, enabling engineers to leverage historical optimization data across different blockchain deployments.

A key observation is that performance can be assessed externally via benchmarking. Hyperledger Caliper provides a standardized framework for generating controlled workloads and reporting metrics such as transactions per second (TPS) and latency statistics~\cite{kaushal2024exploring}. When combined with automated deployment, the end-to-end benchmarking procedure can be viewed as a black-box function that maps a configuration vector to measured performance: given a candidate configuration, the network is instantiated, exercised under a fixed workload, and a scalar metric is returned~\cite{jones1998ego}. This mapping is not available in closed form and is affected by measurement noise due to system nondeterminism and cloud variability. Consequently, the tuning task naturally fits a noisy black-box optimization formulation, where each evaluation is costly and the objective must be optimized with a limited number of benchmark runs~\cite{fabric_docs_performance}.

In this work, we investigate Bayesian optimization (BO) as a sample-efficient approach for optimizing expensive black-box objectives~\cite{snoek2012bo,frazier2018tutorialbo}. BO employs a probabilistic surrogate model to capture uncertainty in the objective and an acquisition function to guide the selection of new configurations, balancing exploration and exploitation. However, standard BO with global surrogate models often becomes ineffective in high-dimensional spaces, since sample complexity increases rapidly with dimension and surrogate fitting/acquisition optimization become unreliable under limited evaluation budgets, motivating high-dimensional BO methods such as random embeddings and local/trust-region variants~\cite{eriksson2021saasbo,eriksson2019turbo}. To address this, we evaluate BO augmented with dimensionality-reduction (DR) techniques, enabling search in lower-dimensional representations while still selecting configurations in the original space~\cite{wang2013rembo}.

This paper makes the following contributions:
\begin{itemize}
    \item We design and implement an end-to-end closed-loop testbed that automatically deploys HLF configurations and benchmarks them using Hyperledger Caliper.
    \item We formulate HLF throughput tuning as a noisy black-box optimization problem over a high-dimensional configuration space derived from HLF configuration files.
    \item We empirically evaluate BO with multiple acquisition functions and DR strategies in a cloud environment, and report measurable throughput improvements for several method combinations under the tested workload.
\end{itemize}

Under the evaluated workload, the best-performing method achieves a 12\% TPS improvement relative to the first evaluated configuration.

Our work sits at the intersection of several active research areas in blockchain and distributed systems. The automated performance tuning of Hyperledger Fabric addresses scalability challenges in permissioned blockchain platforms. The Caliper-in-the-loop framework contributes to benchmarking and simulation tools for blockchain systems. By applying Bayesian optimization and dimensionality reduction techniques, we demonstrate how machine learning methods can be effectively integrated into blockchain configuration and management. These contributions are particularly relevant for enterprise deployments where efficient configuration optimization is critical.

The remainder of the paper is organized as follows. Section~\ref{sec:background} reviews background and related work on Hyperledger Fabric, Hyperledger Caliper benchmarking, black-box optimization, BO, and high-dimensional optimization via dimensionality reduction. Section~\ref{sec:problem} formulates throughput tuning as a noisy black-box optimization problem. Section~\ref{sec:method} describes the proposed Caliper-in-the-loop tuning pipeline, the evaluated network/workload, and the optimization methods and protocol. 
Section~\ref{sec:exp_setup} details the experimental environment, infrastructure, performance metrics, and benchmarking configuration.
Section~\ref{sec:results} presents experimental results and compares optimization strategies. Finally, Sections~\ref{sec:discussion} and~\ref{sec:conclusion} discuss limitations and conclude the paper with directions for future work.



\section{Background and Related Work}
\label{sec:background}

\subsection{Hyperledger Fabric Performance Fundamentals}
\label{subsec:fabric_fundamentals}

HLF is a permissioned blockchain platform designed around modular components for membership, endorsement, ordering, and state management. A defining feature is its \emph{execute--order--validate} transaction pipeline: transaction proposals are first simulated and endorsed by peers, then ordered into blocks by an ordering service, and finally validated and committed by peers. This separation improves flexibility but also increases the number of configuration degrees of freedom, since performance depends jointly on endorsement policies, block cutting parameters, ordering settings, peer execution/validation behavior, and state database characteristics. Prior work has shown that Fabric throughput and latency are sensitive to both architectural choices (e.g., ordering and validation phases) and configuration parameters such as block size, batching timeouts, endorsement policy complexity, and the choice of state database~\cite{baliga2018fabric_char}. These studies motivate automated configuration tuning: the configuration space is large, interactions are non-linear, and the performance surface can be noisy under realistic deployments.

\subsection{Hyperledger Caliper Benchmarking}
\label{subsec:caliper_background}

Hyperledger Caliper is a blockchain benchmarking framework that executes controlled workloads against a target network using platform-specific adapters and produces standardized reports containing throughput and latency metrics. Caliper separates the \emph{network configuration} (how to connect to the system under test) from the \emph{benchmark configuration} (round structure, rate control, workload module, and duration), enabling repeatable experiments across multiple runs and systems. In the work ~\cite{caliper_architecture}, Caliper plays the role of an external measurement oracle: each candidate Fabric configuration is deployed, exercised under a fixed Caliper workload definition, and mapped to a scalar objective (TPS). This framing is valuable because it avoids assumptions about analytic models of Fabric performance and aligns with how practitioners evaluate performance in real deployments.

\subsection{Black-box Optimization and Bayesian Optimization}
\label{subsec:bo_background}

When a system evaluation is expensive, noisy, and does not provide gradients, optimization is naturally formulated as \emph{black-box optimization}. BO is a prominent sample-efficient approach in this setting. BO maintains a probabilistic surrogate model of the objective function and selects new evaluation points via an acquisition function that trades off exploration (reducing uncertainty) and exploitation (improving the best observed value). Gaussian-process (GP) surrogates are commonly used due to their expressive nonparametric prior and closed-form posterior updates under Gaussian noise assumptions. Acquisition functions such as upper confidence bound (UCB) have theoretical guarantees under GP priors and have been studied in the noisy bandit setting. In the context of blockchain benchmarking, these properties are attractive because each evaluation involves non-trivial deployment and workload execution time, and observations can vary due to nondeterminism and infrastructure fluctuations~\cite{srinivas2010gpucb}.

\subsection{Dimensionality Reduction for High-dimensional BO}
\label{subsec:hdbo_background}

A key practical challenge is scaling BO to high-dimensional configuration spaces. As dimensionality increases, surrogate modeling becomes harder with limited evaluation budgets, and acquisition optimization becomes less reliable. One influential approach is to exploit \emph{effective} (intrinsic) low dimensionality through embeddings. Random embedding Bayesian optimization (REMBO) optimizes in a low-dimensional latent space and maps candidates back into the original domain, enabling BO-style search even when the nominal dimension is very large. Beyond BO-specific techniques, broader systems research has explored automated tuning of Fabric using learning-based approaches. For example, Athena uses deep reinforcement learning to recommend configurations that improve throughput and latency in permissioned blockchain deployments. Recent work also studies “self-driving” tuning directions for blockchains under realistic constraints~\cite{chacko2024blockchainlearn}. Relative to these approaches, our study focuses on BO combined with dimensionality-reduction strategies under a strictly \emph{Caliper-in-the-loop} evaluation protocol, emphasizing sample efficiency and controlled comparison across BO variants in a very high-dimensional configuration space.

\section{Problem Statement}\label{sec:problem}

The objective of this work is to automatically tune the performance of a HLF deployment by selecting configuration parameters that maximize throughput under a fixed benchmark workload. Let $x \in \mathcal{X}$ denote a configuration vector, where $\mathcal{X}$ is a bounded, mixed-type configuration domain derived from HLF configuration files. In our setting, the configuration space is high-dimensional, with $d = 317$ parameters (i.e., $x \in \mathbb{R}^d$ after encoding), and includes numerical, categorical, and boolean variables subject to feasibility constraints (e.g., valid ranges, inter-parameter dependencies, and syntactically valid configuration files). Each candidate $x$ induces a concrete HLF network deployment (peer/orderer settings, channel parameters, and related system options).

We evaluate a configuration $x$ by deploying HLF using $x$ and executing a standardized workload using Hyperledger Caliper. The benchmarking procedure returns a throughput metric, denoted by $f(x)$ (TPS). The true mapping $f:\mathcal{X}\rightarrow \mathbb{R}$ is unknown and does not admit an analytic form; it can only be observed through expensive experimentation. Moreover, benchmark outcomes are affected by stochasticity stemming from system nondeterminism and cloud variability. Therefore, each evaluation yields a noisy observation
\begin{equation}
    y(x) = f(x) + \varepsilon,
\end{equation}
where $\varepsilon$ captures measurement noise and runtime fluctuations.

The tuning task can thus be stated as a noisy black-box optimization problem:
\begin{equation}
    x^\star = \arg\max_{x \in \mathcal{X}} f(x),
\end{equation}
subject to feasibility constraints that ensure that $x$ produces a valid deployable HLF configuration. In practice, we have access only to noisy observations $y(x)$, and each evaluation incurs a significant cost (deployment and benchmarking time). Hence, the goal is to identify a high-performing configuration $x^\star$ while minimizing the number of benchmark evaluations.

To account for run-to-run variability, we normalize throughput measurements using
a fixed reference configuration $x_{\mathrm{ref}}$ evaluated repeatedly. Let
\begin{equation}
\bar{y}_{\mathrm{ref}} = \frac{1}{K}\sum_{k=1}^{K} y_k(x_{\mathrm{ref}})
\end{equation}
denote the mean TPS over $K$ repeated runs. For any evaluated configuration $x$,
we report the normalized throughput
\begin{equation}
\tilde{y}(x) = \frac{y(x)}{\bar{y}_{\mathrm{ref}}}.
\end{equation}

\section{System and Methodology}
\label{sec:method}

\begin{figure}[h]
\centering
\includegraphics[width=0.9\columnwidth]{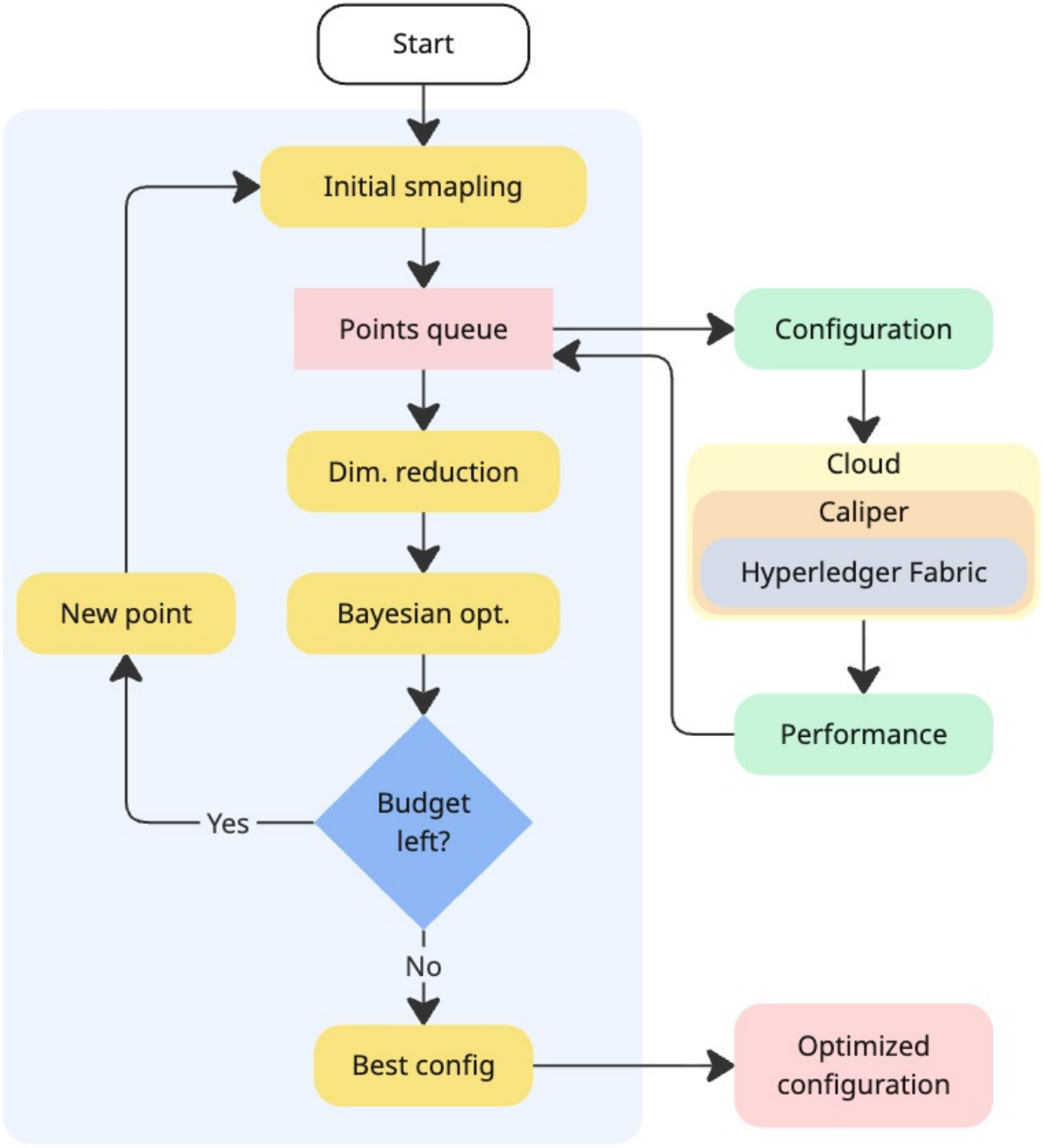}
\caption{Caliper-in-the-Loop closed-loop optimization framework for Hyperledger Fabric performance tuning. The machine learning engine iteratively proposes configurations that are deployed on Yandex Cloud and benchmarked using Hyperledger Caliper and Fabric. Performance metrics feed back into the optimization process until the evaluation budget is exhausted.}
\label{fig:caliper_in_loop}
\end{figure}

This section describes the end-to-end experimental framework used to perform automated HLF tuning. We first present the closed-loop pipeline that integrates configuration generation, network deployment, Hyperledger Caliper benchmarking, and optimizer updates into a single iterative process. Figure~\ref{fig:caliper_in_loop} provides a high-level overview of this Caliper-in-the-Loop framework. We then specify the evaluated network topology and workload, summarize the BO variants considered (acquisition functions and dimensionality-reduction strategies), and conclude with the experimental protocol used to ensure a controlled comparison across methods.

\subsection{Closed-loop Tuning Pipeline}
\label{subsec:closedloop}

We implement an end-to-end closed-loop tuning pipeline that automates the iterative cycle of configuration selection, network deployment, benchmarking, and optimizer update. The pipeline operationalizes the black-box objective introduced in Section~\ref{sec:problem}: each candidate configuration is treated as an input, and the resulting throughput measured under a fixed workload is treated as a (noisy) output observation. By construction, the internal mapping between configuration parameters and performance is not assumed to be known, differentiable, or analytically tractable.

At a high level, the pipeline consumes a configuration vector $x \in \mathcal{X}$ and produces a benchmark observation $y(x)$, where $y(x)$ is the TPS value extracted from Hyperledger Caliper reports. The system is designed to minimize manual intervention because each evaluation requires substantial time for deployment and benchmarking. In addition, the pipeline enforces consistent experimental conditions across iterations, thereby supporting fair comparisons between configurations and across optimization strategies.

Each tuning iteration proceeds through the following steps:
\begin{enumerate}
    \item \textbf{Configuration materialization.} The candidate vector $x$ is decoded into deployable HLF configuration artifacts by populating a set of parameterized templates. This step enforces basic type and range constraints (e.g., boolean, categorical, and bounded numerical values) before deployment.
    \item \textbf{Network instantiation.} An HLF network is created from the generated artifacts. The pipeline initializes peers and the ordering service, creates the required channel(s), and deploys the target chaincode. The iteration proceeds only after the network reaches a ready state.
    \item \textbf{Benchmark execution.} Hyperledger Caliper executes a predefined workload against the running network via the Fabric adapter. The workload definition and benchmark duration are held constant across iterations to ensure that performance differences primarily reflect configuration changes rather than workload variation.
    \item \textbf{Metric extraction.} After each benchmark run, Caliper generates a report containing throughput and latency statistics. The pipeline parses the report and extracts the objective value $y(x)$ used for optimization (TPS in our experiments).
    \item \textbf{Optimizer update and proposal.} The observation pair $(x, y(x))$ is added to the history of evaluated configurations. The optimizer updates its internal state (e.g., surrogate model and any dimensionality-reduction representation) and produces the next candidate configuration for evaluation.
\end{enumerate}

The pipeline accounts for practical feasibility constraints that arise in HLF deployments. Certain parameter combinations can lead to invalid or non-starting networks due to implicit interdependencies across components (e.g., inconsistent batching parameters or incompatible service settings). When an iteration fails to deploy or to produce a valid Caliper report, the configuration is marked infeasible and recorded as a failed evaluation; the tuning process then continues according to the method-specific policy.

This closed-loop architecture provides two core benefits for the study. First, it enables sample-efficient optimization by tightly integrating benchmarking results into the candidate-selection mechanism. Second, it supports parallelization: independent optimization trials can be executed on separate machines using the same deployment and benchmarking workflow, facilitating controlled comparisons among optimization strategies under similar operating conditions.

\subsection{Blockchain Network and Workload}
\label{subsec:network_workload}

To evaluate configuration-tuning methods under controlled and repeatable conditions, we benchmark a fixed HLF network topology and a fixed application workload across all optimization iterations. The network configuration is instantiated from a common template, while the tuning process modifies selected parameters within this template according to the candidate configuration vector $x$.

\subsubsection{Network topology}
\label{subsubsec:topology}

The experimental network consists of five organizations: four peer organizations and one ordering organization. Each peer organization hosts one peer node, while the ordering organization hosts the ordering service. Identities and cryptographic material are generated using \texttt{cryptogen}, and the network is bootstrapped using standard Fabric configuration artifacts (e.g., genesis block and channel configuration). This topology is chosen to reflect a representative multi-organization permissioned setting while remaining sufficiently lightweight to support repeated deployments during optimization.

The network follows Fabric's \emph{Execute--Order--Validate} transaction pipeline. Endorsing peers simulate transaction proposals and produce endorsements; the ordering service sequences endorsed transactions into blocks; and peers validate and commit ordered blocks to the ledger. For state management, the deployed network uses a consistent world-state backend across all runs, ensuring that observed differences in throughput primarily reflect tuned parameters rather than changes in storage configuration.

\subsubsection{Chaincode and application workload}
\label{subsubsec:workload}

We benchmark the network using a fixed chaincode and workload profile throughout the study. Specifically, we use the \texttt{FabCar} chaincode as a representative key-value application frequently used in Fabric performance demonstrations and benchmarking examples. The workload is executed via Hyperledger Caliper, which submits transactions at controlled rates and reports performance metrics including throughput (TPS) and latency statistics.

To ensure comparability across candidate configurations, the benchmark definition is held constant across all evaluations. Each benchmark run follows the same procedure: the network is deployed from the current configuration, Caliper connects to the running network through the Fabric adapter, transactions are submitted according to the predefined workload module, and performance metrics are collected from the resulting Caliper report. This design isolates the effect of configuration changes on observed performance and provides a consistent basis for comparing optimization methods.

\subsubsection{Metric of interest}
\label{subsubsec:metric}

The primary optimization target in this paper is throughput, measured as TPS as reported by Caliper. TPS is widely used as a first-order indicator of system capacity in permissioned blockchains and is directly comparable across tuning iterations when the workload specification is fixed. Where available, we also record auxiliary metrics (e.g., latency summaries) for diagnostic analysis, but the optimization objective is defined solely in terms of TPS to keep the black-box tuning problem well specified and to enable direct comparison across optimization strategies.

\subsection{Optimization Methods (AF and DR)}
\label{subsec:opt_methods}

We study automated HLF tuning using BO under the noisy black-box setting defined in Section~\ref{sec:problem}. BO is attractive in this context because benchmark evaluations are expensive and the objective function does not admit gradients or an analytic form. At each iteration, BO maintains a probabilistic surrogate model of the objective and uses an acquisition function to select the next configuration to evaluate. The surrogate captures both predicted performance and epistemic uncertainty, while the acquisition function trades off exploration of poorly understood regions and exploitation of regions predicted to yield high throughput.

A central challenge for applying BO to HLF tuning is the dimensionality of the configuration space. In our setting the configuration vector is high-dimensional ($d=317$ after encoding), and standard BO can become inefficient due to the curse of dimensionality and the difficulty of fitting accurate surrogate models with limited data. We therefore evaluate BO augmented with dimensionality-reduction (DR) strategies that aim to identify a lower-dimensional structure in the search space, enabling more effective modeling and candidate selection while still producing valid configurations in the original space.

\subsubsection{Acquisition functions}
\label{subsubsec:acq}

We consider four acquisition functions that are commonly used in BO and that represent different exploration, exploitation behaviors:
\begin{itemize}
    \item \textbf{Expected Improvement (EI):} selects configurations that maximize the expected improvement over the best observed objective value under the surrogate distribution.
    \item \textbf{Maximum Probability of Improvement (MPI):} selects configurations that maximize the probability of outperforming the current best observation by a specified margin.
    \item \textbf{Upper Confidence Bound (UCB):} selects configurations that maximize a weighted combination of the surrogate mean and uncertainty, with the weight controlling the exploration level.
    \item \textbf{DYCORS:} a dynamic coordinate-search strategy that perturbs a subset of dimensions at each iteration, encouraging local exploration while remaining scalable in high-dimensional settings.
\end{itemize}

\subsubsection{Dimensionality-reduction strategies}
\label{subsubsec:dr}

To mitigate the impact of high dimensionality, we evaluate four DR strategies that compress the configuration representation used by the optimizer:
\begin{itemize}
    \item \textbf{PCA:} applies principal component analysis to identify a low-dimensional linear subspace capturing the dominant variance in the observed configurations.
    \item \textbf{REMBO:} optimizes the objective in a low-dimensional embedding and maps candidate points back to the original space via a random projection, targeting effective dimensionality rather than nominal dimensionality.
    \item \textbf{SA:} a sensitivity-based reduction that ranks parameters by their observed influence on throughput within the collected samples and focuses subsequent search on the highest-ranked subset.
    \item \textbf{SHAP:} a feature-attribution-based ranking computed from a fitted predictive model over the observed data, used to prioritize parameters for subsequent search.

\end{itemize}

\subsubsection{Method combinations}
\label{subsubsec:combos}

Our evaluation considers all combinations of the four acquisition functions and four DR strategies, yielding $4 \times 4 = 16$ BO variants. In addition, we include a \textbf{random search} baseline that samples configurations uniformly from the feasible domain. Each method operates under the same evaluation budget and workload definition, and each method produces a sequential stream of candidate configurations based on the observations collected during its trial. This design enables controlled comparison of acquisition-function behavior and DR choice under identical benchmarking conditions.

\subsection{Experimental Protocol}
\label{subsec:protocol}

Our experimental protocol is designed to enable controlled comparison of optimization strategies under identical benchmarking conditions, while accounting for the high cost and noise of blockchain performance measurements. All methods are evaluated using the same network topology, the same Caliper workload profile, and the same computational environment (Section~\ref{subsec:network_workload}).

\subsubsection{Initialization and evaluation budget}
\label{subsubsec:budget}

Each optimization trial follows a fixed evaluation budget of $B=300$ benchmark evaluations. Since a single evaluation includes configuration materialization, network deployment, and benchmark execution, the wall-clock cost per evaluation is non-trivial (on the order of minutes). To improve initial coverage of the search space and reduce cold-start effects, we use an initial space-filling design based on Latin Hypercube Sampling (LHS). The LHS phase produces an initial set of configurations that are evaluated before the sequential BO loop proceeds.

After initialization, each method performs sequential candidate selection: at iteration $t$, the method proposes a single new configuration $x_t$, the pipeline evaluates it to obtain $y(x_t)$, and the observation is incorporated into the method state before proposing $x_{t+1}$. This sequential protocol ensures that all methods operate under the same information constraints (i.e., candidates are selected only based on observations available at that point in the trial).

\subsubsection{Parallel execution of trials}
\label{subsubsec:parallel}

To compare methods under similar conditions and within practical time limits, we execute trials in parallel across multiple machines. Specifically, we evaluate the 16 BO variants (Section~\ref{subsec:opt_methods}) and one random-search baseline concurrently, with each method assigned a dedicated virtual machine. Parallel execution reduces total experiment time while preserving within-trial sequential decision making (i.e., each method remains sequential internally, but different methods run concurrently).

\subsubsection{Measurement normalization and noise handling}
\label{subsubsec:noise}

Benchmark measurements exhibit run-to-run variability due to system nondeterminism
and cloud fluctuations. To enable fair cross-method comparison, we normalize
throughput using a fixed reference configuration $x_{\mathrm{ref}}$. We evaluate
$x_{\mathrm{ref}}$ repeatedly ($K$ runs) and compute
\begin{equation}
\bar{y}_{\mathrm{ref}} = \frac{1}{K}\sum_{k=1}^{K} y_k(x_{\mathrm{ref}}).
\end{equation}
For any evaluated configuration $x$, we report
\begin{equation}
\tilde{y}(x) = \frac{y(x)}{\bar{y}_{\mathrm{ref}}}.
\end{equation}
In our experiments, $x_{\mathrm{ref}}$ is chosen as the first evaluated
configuration and $K=10$.

\subsubsection{Handling infeasible configurations and failed runs}
\label{subsubsec:failures}

Certain parameter combinations can produce infeasible deployments (e.g., invalid configurations or networks that fail to start) or unsuccessful benchmark runs (e.g., inability to obtain a valid Caliper report). When such failures occur, the corresponding candidate is recorded as a failed evaluation. For fairness, all methods are subject to the same feasibility checks and failure-handling rules. Failed evaluations do not contribute a valid throughput observation; the trial continues to the next iteration according to the method's proposal mechanism.

\subsubsection{Reproducibility and logging}
\label{subsubsec:repro}

For each evaluation, the system logs: the proposed configuration vector, the instantiated configuration artifacts, deployment status, benchmark configuration, and the resulting Caliper outputs. This enables post-hoc inspection of best-performing configurations and supports reproducibility of reported results. Where applicable, pseudo-random components (e.g., initialization sampling) are controlled through explicit random seeds.

\section{Experimental Setup}
\label{sec:exp_setup}

\subsection{Cloud Infrastructure}
\label{subsec:cloud}

All experiments were executed in a public cloud environment to support repeated redeployments and parallel evaluation of multiple optimization strategies. We provisioned a pool of virtual machines (VMs) in Yandex Cloud and assigned each optimization trial to a dedicated VM to reduce cross-trial interference and preserve sequential decision making within each optimizer. Yandex Compute Cloud provides configurable VM platforms and vCPU performance levels, enabling controlled specification of compute resources across trials~\cite{yandex_vm_platforms,yandex_vcpu_perf}.

Each VM was configured with a fixed CPU/RAM profile and a persistent SSD-backed boot disk, and the same deployment and benchmarking toolchain was used across all instances. HLF components were deployed and benchmarked through an automated pipeline (Section~\ref{subsec:closedloop}), and performance metrics were collected from Hyperledger Caliper reports~\cite{fabric_docs_txflow,caliper_repo,caliper_arch_doc}. Fixing the VM resource profile and execution workflow across trials ensures that observed throughput differences are attributable primarily to configuration choices and optimizer behavior rather than heterogeneous infrastructure.

\subsection{Metrics}
\label{subsec:metrics}

We evaluate each candidate configuration using performance indicators reported by Hyperledger Caliper. Caliper executes a specified workload against the system under test and produces an aggregate performance report, including throughput and latency statistics~\cite{caliper_site,caliper_repo,lf_caliper_blog}. In this study, the optimization objective is throughput, while additional indicators are recorded to diagnose failures and interpret performance trends.

The primary metric is \textbf{throughput}, measured in \textbf{TPS}. We use the TPS value reported by Caliper as the objective observation $y(x)$ for a configuration $x$. Caliper reports both the \emph{send rate} (the rate at which the benchmark driver issues transactions) and the \emph{achieved rate} (the effective throughput observed as successful transactions are processed)~\cite{ibm_caliper_metrics,caliper_repo}.

To characterize responsiveness, we record \textbf{transaction latency}, defined as the elapsed time from client submission until the transaction is processed and committed, as captured by Caliper's latency measurements~\cite{caliper_site,ibm_caliper_metrics,perf_metrics_whitepaper}. We retain summary latency statistics (minimum, maximum, and average) and, where available, percentile values reported by Caliper.

To assess reliability under load, we record \textbf{success and failure counts} for submitted transactions and compute a \textbf{success rate} from Caliper's reported counts~\cite{caliper_site,caliper_repo}. These indicators allow us to distinguish genuine throughput improvements from cases where throughput changes are driven by increased failure rates.

Finally, Caliper supports \textbf{resource consumption} metrics (e.g., CPU, memory, and network I/O) through metric adapters and monitoring integrations~\cite{caliper_site,caliper_repo}. When such measurements are available in the benchmark report for a given run, we store them as auxiliary diagnostics; however, the optimization objective in this paper is defined solely in terms of TPS to enable direct comparison across methods.

\subsection{Benchmark Configuration}
\label{subsec:bench_config}

All evaluations use a fixed Hyperledger Caliper benchmark specification to ensure that observed performance differences are attributable to HLF configuration changes rather than workload variation. Caliper separates (i) the \emph{benchmark configuration}, which defines how an experiment is executed, from (ii) the \emph{network configuration}, which specifies how Caliper connects to the system under test~\cite{caliper_site,caliper_architecture}.

The benchmark configuration file (YAML or JSON) defines a sequence of \emph{rounds} and their execution characteristics~\cite{caliper_bench_config}. Each round specifies (1) its execution length, controlled either by \texttt{txDuration} (time-based execution) or \texttt{txNumber} (count-based execution); (2) a \texttt{rateControl} policy that determines the transaction submission behavior (e.g., fixed-rate controllers); and (3) a \texttt{workload} module responsible for generating transaction content and request parameters~\cite{caliper_bench_config,ibm_caliper_tutorial}. Caliper then executes the defined rounds against the deployed network via the appropriate blockchain adapter and produces an aggregate report containing throughput and latency statistics~\cite{caliper_site,caliper_architecture}.

Across all optimization iterations, we keep the benchmark definition constant: the set of rounds, the \texttt{rateControl} policy, and the workload-module parameters do not change between evaluated configurations. Consequently, the only changing element between evaluations is the HLF deployment configuration induced by the candidate vector $x$, while the benchmark workload and submission policy remain fixed for the duration of the study.

\section{Results}
\label{sec:results}

We report results for 16 variants of BO (4 acquisition functions $\times$ 4 DR strategies) and a random-search baseline, each executed under the same workload, evaluation budget, and infrastructure profile. Unless stated otherwise, improvements are reported relative to the reference
configuration $x_{\mathrm{ref}}$ (the first evaluated configuration in each trial),
consistent with the normalization in Section~\ref{subsubsec:noise}.

\begin{figure}[h]
\centering
\includegraphics[width=\columnwidth]{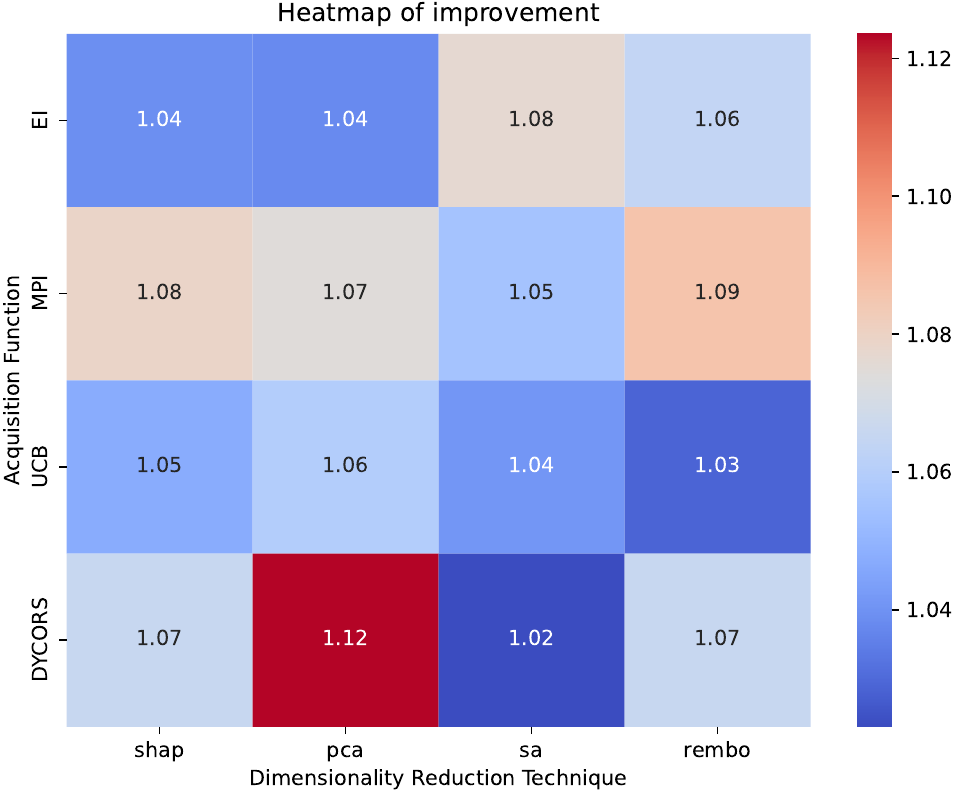}
\caption{Best achieved throughput improvement factor by method (higher is better), reported relative to the first evaluated configuration within each trial.}
\label{fig:improvement_heatmap}
\end{figure}

Figure~\ref{fig:improvement_heatmap} summarizes the best improvement factor achieved by each acquisition-function/DR combination. Several combinations yield non-trivial gains, but performance depends on the specific pairing of acquisition function and DR method; no single acquisition function dominates uniformly across DR choices. The strongest cell in the grid is DYCORS--PCA with an improvement factor of 1.12 (corresponding to a 12\% gain), followed by MPI--REMBO at 1.09 (9\%). Multiple other combinations achieve moderate gains in the 1.04--1.08 range, while the weakest observed improvement factors are close to 1.02--1.03.

\begin{figure}[h]
\centering
\includegraphics[width=\columnwidth]{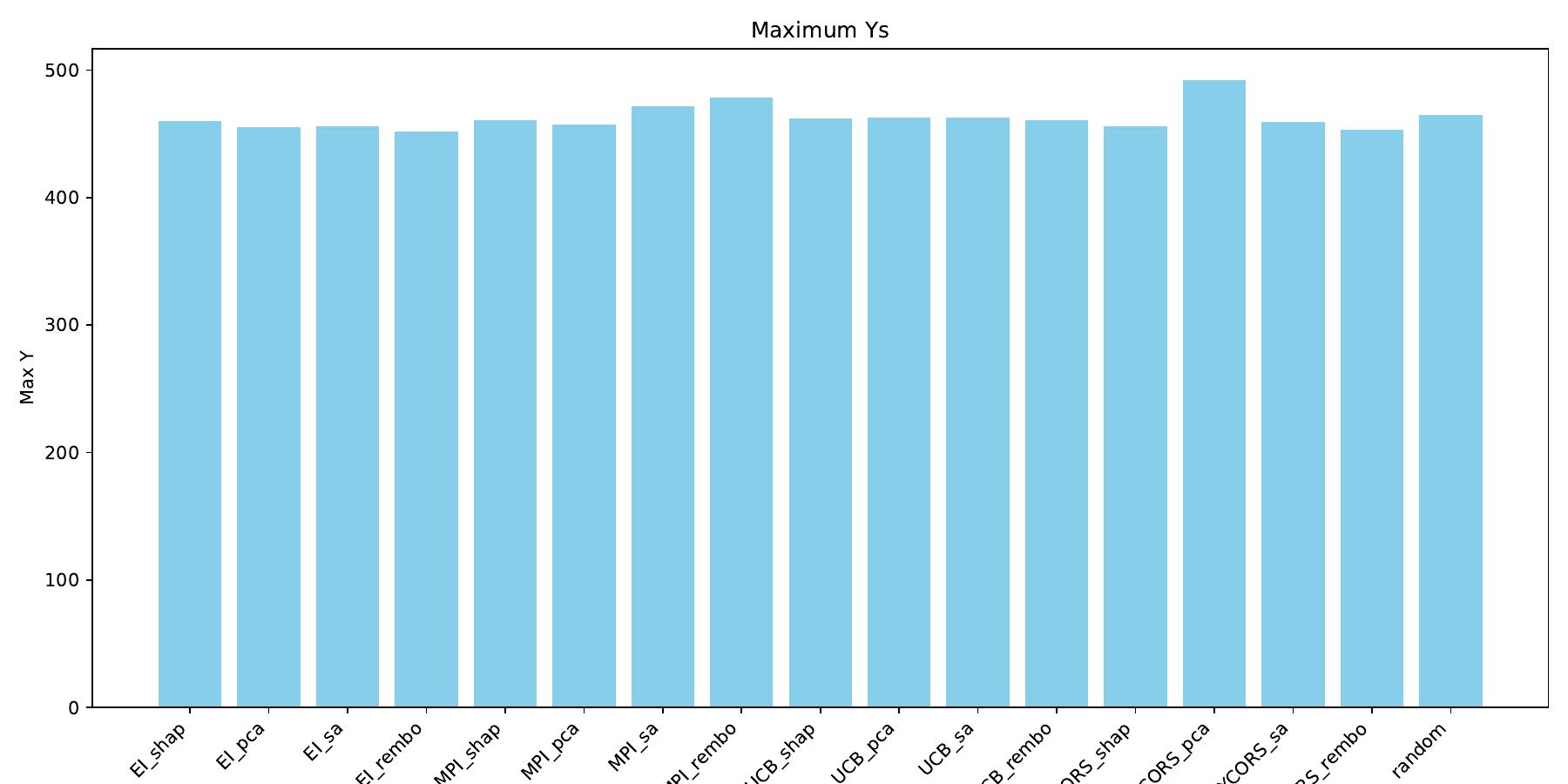}
\caption{Maximum TPS observed in each trial (absolute throughput). Bars correspond to AF--DR pairs and the random-search baseline.}
\label{fig:best_values}
\end{figure}

Figure~\ref{fig:best_values} reports the maximum (absolute) TPS observed in each trial. Consistent with the improvement heatmap, DYCORS--PCA attains the highest maximum TPS among the evaluated methods, and MPI--REMBO is among the strongest performers. The bar chart also highlights that many method variants end within a relatively narrow band of maxima under the tested workload, suggesting that (i) the workload and infrastructure impose a practical performance envelope, and (ii) gains are primarily achieved by identifying configurations near the upper end of this envelope rather than by shifting the envelope dramatically.

\begin{figure}[h]
\centering
\includegraphics[width=\columnwidth]{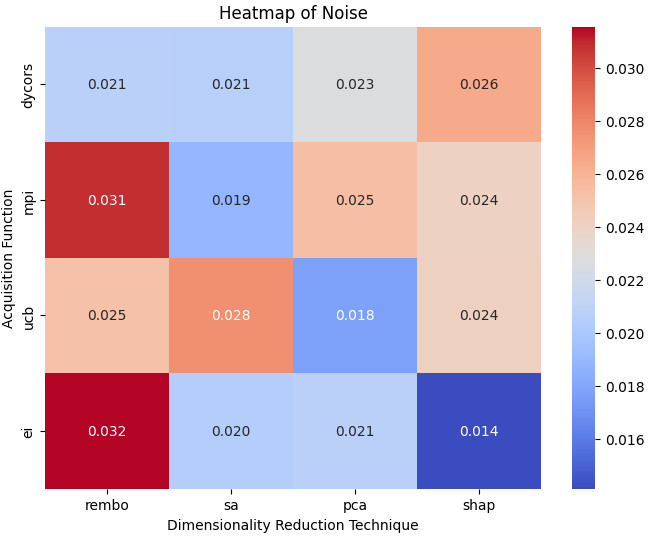}
\caption{Noise scores across AF--DR trials (lower indicates more stable measurements). Values are computed from repeated evaluations of the reference configuration used for normalization (Section~\ref{subsubsec:noise}).}
\label{fig:noise_heatmap}
\end{figure}

Because HLF benchmarking is subject to run-to-run variability, we quantify measurement variability per trial and normalize results using a fixed reference configuration evaluated repeatedly (Section~\ref{subsubsec:noise}). Figure~\ref{fig:noise_heatmap} shows the resulting noise scores for each AF--DR pairing. Across trials, the noise scores lie in a narrow range (approximately 0.014--0.032), indicating a moderately stable but non-negligibly noisy measurement environment. Variability is not uniform across methods: some combinations exhibit systematically higher noise than others, motivating the use of normalization and caution when interpreting small performance differences between closely ranked methods. Importantly, the largest reported improvements (e.g., DYCORS--PCA at 12\% and MPI--REMBO at 9\%) exceed the observed noise range, making it unlikely that these gains are solely attributable to measurement fluctuations under this protocol.

\begin{figure}[h]
\centering
\includegraphics[width=\columnwidth]{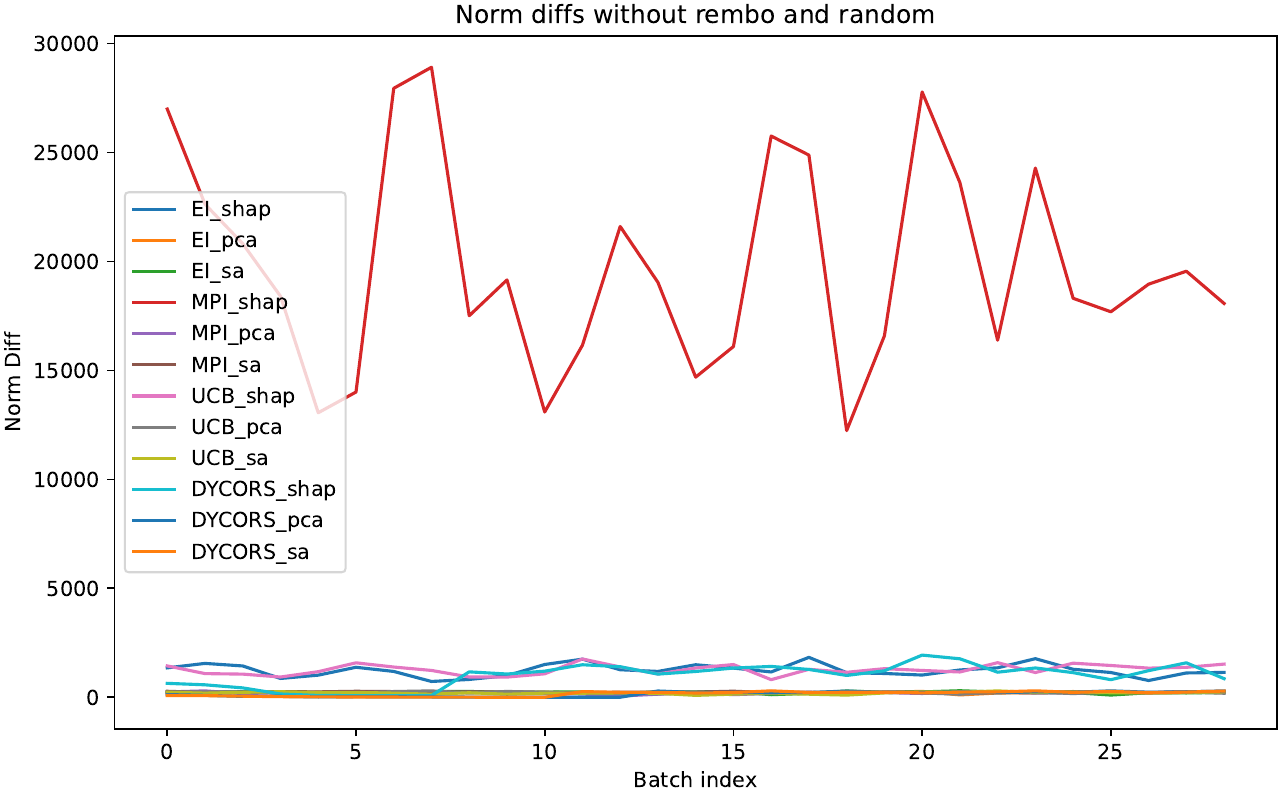}
\caption{Batched norm differences between successive proposed configurations for non-REMBO, non-random trials (lower indicates more local search steps).}
\label{fig:normdiffs}
\end{figure}

To provide qualitative insight into optimizer behavior in the 317-dimensional encoded space, we examine the magnitude of changes between successive proposed configurations. Figure~\ref{fig:normdiffs} plots the (batched) norm difference between consecutive configuration vectors for all non-REMBO and non-random trials, avoiding scale domination by methods that generate very large jumps. Most methods exhibit relatively small successive changes, consistent with localized search behavior once promising regions are identified. In contrast, at least one method trace shows substantially larger jumps across many batches, reflecting more aggressive exploration. Notably, larger exploration steps do not necessarily translate to the highest throughput gains (e.g., some exploratory behaviors achieve moderate improvements but are outperformed by DYCORS--PCA in Figure~\ref{fig:improvement_heatmap}), highlighting the trade-off between exploration and exploitation in this noisy, high-dimensional setting.

\section{Discussion and Limitations}
\label{sec:discussion}

Our results indicate that Caliper-in-the-loop black-box optimization can identify HLF configurations that improve throughput under a fixed workload and infrastructure profile. The strongest BO+DR combinations achieve clear gains within their trials, suggesting that sample-efficient search is a practical alternative to manual trial-and-error when the configuration space is large and parameter interactions are difficult to anticipate. However, performance is not uniform across acquisition-function/DR pairings, emphasizing that the choice of optimizer and dimensionality-reduction strategy is itself a meaningful design decision in high-dimensional tuning.

The main limitations concern generality and measurement uncertainty. The reported gains are conditional on the evaluated topology, chaincode/workload, and VM profile; different endorsement policies, state database choices, and client concurrency can shift bottlenecks and alter which parameters matter. Benchmark outcomes also vary across runs in cloud environments; while normalization mitigates drift, small differences between closely ranked methods may remain within the variability band, and results can be sensitive to the baseline and evaluation protocol. Finally, the mixed-type configuration space includes feasibility constraints, and invalid configurations can consume evaluation budget; incorporating constraint-aware modeling and confirmation runs of the best configurations would strengthen robustness.

\section{Conclusion}
\label{sec:conclusion}

This paper studied automated throughput tuning of HLF by treating Caliper-based benchmarking as a noisy black-box objective. We implemented an end-to-end closed-loop pipeline that deploys candidate configurations, benchmarks them with Hyperledger Caliper, and updates an optimizer to propose the next configuration in a 317-dimensional search space derived from Fabric configuration artifacts. In a cloud testbed, we evaluated 16 BO variants combining four acquisition functions with four dimensionality-reduction strategies, alongside a random-search baseline. Under the tested workload, the best method (DYCORS--PCA) achieved a 12\% throughput improvement relative to the first evaluated configuration, and MPI--REMBO achieved a 9\% improvement, demonstrating that BO combined with DR can be effective for high-dimensional Fabric tuning. Future work will extend the evaluation to additional workloads and multi-objective tuning criteria that jointly capture throughput, latency, and reliability.


\bibliographystyle{IEEEtran}
\bibliography{bibliography}

\end{document}